\documentclass[11pt,a4paper]{article}

\usepackage[margin=1in]{geometry}
\usepackage[utf8]{inputenc}
\usepackage[T1]{fontenc}
\usepackage{lmodern}
\usepackage{microtype}
\usepackage{graphicx}
\usepackage{booktabs}
\usepackage{array}
\usepackage{xcolor}
\usepackage{hyperref}
\usepackage{url}
\usepackage{amsmath}
\usepackage{amssymb}
\usepackage{multirow}
\usepackage{caption}

\hypersetup{
    colorlinks=true,
    linkcolor=blue!60!black,
    citecolor=blue!60!black,
    urlcolor=blue!60!black,
}

\title{Empirical Study of Pop and Jazz Mix Ratios \\ for Genre-Adaptive Chord Generation}
\author{Jinju Lee \\ \small{PearlLeeStudio} \\ \small{\texttt{pearl1379@gmail.com}}}
\date{May 2026}

\begin{document}
\maketitle

\begin{abstract}
Chord progression generation is practically important but understudied. Most large-scale symbolic music systems target melody, multi-track arrangement, or audio synthesis, and chord-only models tend to be relegated to conditioning components inside larger pipelines. This paper treats chord generation as a standalone task and addresses a question that arises whenever such a model is adapted across genres: how much old-domain data must be retained during fine-tuning to acquire a new domain without forgetting the old? I study jazz fine-tuning starting from a pop-pretrained 25M-parameter Music Transformer (84.21\% top-1 next-token accuracy on a held-out pop test set). The available jazz corpus is more than two orders of magnitude smaller than the pop corpus, so every fine-tune run uses all 1{,}513 jazz training sequences. The swept variable is the volume of pop ``rehearsal'' data mixed alongside, taking values in $\{0, 1\text{K}, 2.5\text{K}, 5\text{K}, 10\text{K}\}$. At the tabulated best epochs every fine-tuned run gains 7 to 9 points of jazz top-1 (the \emph{released} F1 checkpoint coincides with the Phase~0 baseline, as the Correction below explains), and pop top-1 falls by 2.11 points under jazz-only fine-tuning but by at most 1.19 with rehearsal. Top-1 also scores the three header tokens of each sequence (key, time signature, genre marker), and two of them shift whole runs: the jazz-majority runs predict the jazz marker even for pop songs, and most test songs carry a placeholder C-major key that a checkpoint predicts for every such song or for none, as the best epochs of F1 and F3 happen to do. Scored without the header tokens, every run gains 7.2 to 7.8 jazz points, jazz-only fine-tuning loses 1.48 pop points, and the four rehearsal volumes (1K to 10K, $0.66\times$ to $6.6\times$ the jazz volume) lose pop at nearly one rate, within 0.11 points of each other at every epoch (the 10K run losing the least) and 0.73 to 0.90 points below baseline by their last epoch. The smallest volume tested therefore already gives nearly all of the measured protection, and no volume stops a slow decline. A complementary observation: the middle of the sweep (F3, 2.5K mix) is not always the perceptually preferred run. The pop-leaning (10K) and jazz-leaning (1K) endpoints carry more committed stylistic identities that the author more often selects as finished output in informal listening. I discuss what this suggests for music co-creation tools but make no perceptual claim, since no formal listening study has been conducted. All six checkpoints and the selection-corrected F1 retrain \texttt{ft-pop80-v2} are released on the HuggingFace Hub at \url{https://huggingface.co/PearlLeeStudio}.
\end{abstract}

\noindent\textbf{Correction (v2).} Tables~4--5 report metrics at each run's \emph{best training epoch} (Section~5.4), computed from the per-epoch CSVs. The checkpoints actually \textbf{released} on the Hub were instead selected by \textbf{minimum mixed-validation loss}, which differs from the best epoch for three of the six runs. Most consequentially, the released F1 (\texttt{ft\_jazz\_pop80}) is \textbf{weight-identical to the Phase~0 pop baseline} (full SHA-256 over all 25{,}841{,}152 parameters\footnote{The SHA-256 covers the 25{,}841{,}152-element serialized state dict. The model has 25{,}661{,}440 unique parameters, the difference being the tied input/output embedding ($351\times512$) stored as two tensors.}: the run resumed with the Phase~0 optimizer and scheduler state, whose learning rate had decayed to 0, so its first fine-tune epoch updated no weights, and the pop-dominated mixed-validation loss was minimized at that unchanged epoch). Therefore: (i)~the released F1 scores jazz 72.86\,/\,pop 84.21 (the Phase~0 row), not the 81.03\,/\,84.60 reported in Table~4; (ii)~the Reproducibility claim that the metrics regenerate end-to-end from the released artifacts holds only for F2 and F5 (the best-epoch weights for F1, F3, F4 were never uploaded); (iii)~the Abstract's ``every fine-tuned run gains 7 to 9 points of jazz top-1'' and Section~7.1's ``F1 is the only run that improves on the Phase~0 pop baseline ($+0.39$)'' (withdrawn in v4) describe the best-epoch \emph{training} metrics, not the released F1 (which, being Phase~0, gains nothing). The mix-ratio sweep conclusions are unaffected, as they rest on the per-epoch training trajectories. Full hash verification appears in arXiv:2606.07334 (Section~4.1). A selection-corrected retrain (jazz-only validation selection) restores a hash-distinct jazz-adapted F1, released as \texttt{ft-pop80-v2} (jazz top-1 75.05 on the expanded 9-source test). The selection rule reproduces across three random seeds in matched-data retrains on that corpus (jazz top-1 $75.76 \pm 0.03$). This v2 also reports the Phase~0 pop figure on the held-out test set ($84.21$, consistent with the fine-tuned rows), which uniformly shifts the Table~5 pop deltas by $0.03$. The conclusions are unchanged. Separately, this revision corrects a transcription error in the v1 Tables~4--5 F2 row, which had inadvertently duplicated F1's epoch-4 metrics. The correct F2 (5K mix) best-epoch values are pop 84.06\,/\,jazz 79.91 (deltas $-0.15\,/\,+7.05$), which leaves every conclusion unchanged.

\noindent\textbf{Revision (v3).} Corrects description errors that affect no table, released artifact, or conclusion: the vocabulary composition in Sections~2.1 and~4.2 (312 chord + 24 key + 5 time-signature + 6 genre + 4 structural tokens: earlier text miscounted the key tokens and misdescribed the genre markers), the Sections~4.1 and~5.2 characterization of the pop pool (deduplication is jazz-side only, and approximately 544K is the 80\% train split of the 680{,}373-song pop pool, not a deduplication result), the Sections~6.2 and~7.2 learning-curve prose (F1--F3 stay within a point (not 0.5) of the pop baseline, worst dips $-0.71$/$-0.82$/$-0.82$; F5 loses 1.4 points in its first updating epoch (epoch~4) and slides to $-2.1$, not ``about two points'' at once; F4 ends $-1.5$; Section~7.1's ``within 0.5 of the jazz peak'' is 0.51), the corpus-size ratio in the Abstract, the sweep-span phrasing in Section~9, and the v2 note's attribution of the multi-seed $75.76 \pm 0.03$ figure (measured on matched-data retrains, whereas the released \texttt{ft-pop80-v2} scores 75.05 on that test).

\noindent\textbf{Revision (v4).} Adds a measurement that changes the rehearsal conclusion, and corrects descriptions. No logged value, table entry, or released artifact changes. (i)~Top-1 scores every target, including the three header tokens that open each sequence (key, time signature, genre marker), and a checkpoint makes nearly the same header prediction for every test sequence (Section~5.4). The jazz-majority runs F4 and F5 predict the jazz marker even for pop songs. Most test songs carry a placeholder C-major key, which twelve-key augmentation leaves a model predicting almost at random, and the tabulated epochs of F1 and F3 are epochs that predict it, which adds about 0.86 pop and 0.70 jazz points. Without the header tokens (new Table~6, derived from the logged values and checked against the released checkpoints), every run gains 7.2 to 7.8 jazz points, F5 loses 1.48 pop points rather than 2.11, and F1 to F4 lose pop at nearly one rate, within 0.11 points of each other at every epoch. The Abstract and Sections~1, 6, 7, 8, and~9 therefore no longer report a recovery to baseline at 2.5K rehearsal samples or a pop gain for F1, and they report at most a small advantage for the 10K volume: every tested volume from 1K gives nearly the same protection, cutting F5's pop loss by about half or more, and none stops a slow decline. (ii)~Every fine-tune resumed from Phase~0 together with its optimizer and scheduler state, so the configured fine-tune learning rate and warmup never applied: the first fine-tune epoch (epoch~3) ran at the inherited rate of 0 and updated no weights, and the later epochs ran a cosine schedule on the Phase~0 base rate of $3 \times 10^{-4}$ (Section~5.3). The Correction above now names this resumption, not warmup, as the reason the released F1 equals Phase~0, and Section~6.2 places F5's first drop at its first updating epoch (epoch~4). (iii)~Section~4 now states how Chordonomicon's genre tags map onto the three markers, that only JazzStandards is deduplicated against JHT, how chords are spelled and normalized (slash-chord basses are dropped, and spellings outside the alias table fall to string heuristics), that the six corpora are split in one pooled shuffle, and how many test songs repeat a training song. Section~5.4 states which songs carry the placeholder key, JHT's flat keys included. Section~2 corrects the chord events per song and what Phase~0's 73\% jazz top-1 consists of, Section~5 states the Phase~0 schedule as applied and the truncation at 256 tokens, the metric is called top-1 next-token accuracy throughout, the Figure~1 tick for the 2.5K run is relabeled and its sweet-spot band removed, the Reproducibility paragraph points to the public code and lists what each model repository holds, and six bibliography entries are corrected.

\section{Introduction}
\label{sec:intro}

Composition is layered. Different traditions privilege different layers as the natural starting point. Singer-songwriters often start from a vocal melody and refine its harmony as the arrangement takes shape. Classical composers more often work from motivic sketches that develop through harmonic context. Producers in electronic and hip-hop traditions begin from rhythm and timbre. Across many of these traditions, \emph{chord-first composition}, which commits to the harmonic skeleton before melody, lyrics, or arrangement details, is a common practice rather than a separate tradition. It is standard for guitar- and piano-based pop and rock songwriters, for jazz musicians playing over fixed ``changes'' in the great American songbook~\cite{steedman1984generative}, and for contemporary commercial idioms like CCM, K-pop, J-pop, and modern country, where a recognizable progression is itself a primary aesthetic object.

Despite this centrality, chord-progression generation has received little independent attention in deep learning music research. The field has emphasized melodic generation~\cite{huang2019music,payne2019musenet,hsiao2021compound}, multi-instrument arrangement~\cite{ens2020mmm,dong2018musegan,thickstun2024anticipatory}, and end-to-end audio synthesis~\cite{copet2023musicgen,agostinelli2023musiclm}. Chord progressions, when they appear, serve as conditioning inputs to melody models~\cite{yang2017midinet,roberts2018hierarchical} or sub-components of full-arrangement systems. Recent surveys~\cite{ji2020comprehensive,briot2020deep} reflect this bias. The reasons are coherent. Melodies are immediately auditable and admit clean sequence-level metrics. Chord quality is sensitive to notation ambiguity (\texttt{Cm7} vs.\ \texttt{C:min7}), to harmonic context that single-token metrics cannot capture, and to genre-specific conventions about voice leading and substitution~\cite{rohrmeier2011generative}. The literature shifted toward tasks where evaluation feels less brittle.

There is a deeper reason to keep the modeling scope at the level of chord symbols rather than reaching toward full-performance reconstruction. Jazz lives mostly outside its recordings. Its core practice is live improvisation. Improvisation has structure, but no model can observe more than a small fraction of the performances that actually exist. Recordings systematically under-sample the practice, and even within the recorded subset, what makes a performance distinctive (the choices a soloist makes against a particular rhythm section on a particular night) is not preserved. The image-generation analogy is instructive. AI can produce images, but it cannot learn the curatorial intent of a gallery: viewing order, dialogue between adjacent works, sight lines a curator designed. That information is private. In principle such structure could be modeled if it were exposed. In practice, the artists whose work would be required have spent generations defending their right to control how it circulates, and it is not obvious that they would or should grant the access. Working at the chord-symbol level respects that boundary. The symbols are an established notation that musicians already share publicly through lead sheets and chord charts. A model trained on chord symbols operates on what is already in circulation, not on what is not.

The contribution is empirical. I study how much rehearsal data is enough in pop-to-jazz chord fine-tuning, sweeping the rehearsal volume from zero to seven times the target-domain volume in five steps while holding jazz fixed. The headline finding: the rehearsal needed is small relative to the pretraining corpus. Once the header tokens that open every sequence are set aside (Section~\ref{sec:method-eval}), a rehearsal volume of two thirds of the jazz training volume protects pop nearly as well as one of seven times that volume, cutting the forgetting of jazz-only fine-tuning by about half or more, but no volume tested stops a slow decline over epochs. A secondary, more tentative finding is that the \emph{aesthetically} preferred outputs in informal listening cluster at the \emph{endpoints} of the sweep rather than in its middle. I discuss what that suggests for symbolic music co-creation tools but make no perceptual claim. The motivating empirical anchor was a specific failure during development of a chord-composition application maintained by the author. An earlier version used a pop pretrain followed by a jazz-only fine-tune and produced output that knowledgeable users called ``technically jazz'' but ``too dense to use.'' The diagnosis was catastrophic forgetting~\cite{mccloskey1989catastrophic,french1999catastrophic,goodfellow2013empirical}. The pop pretrain instilled fluency in commercial harmonic vocabulary. The jazz-only fine-tune rewrote much of it in place. The resulting model drifted into harmonic territory that worked for fluent jazz musicians but not for a wider audience. This regime is exactly what continual learning studies under \emph{rehearsal} or \emph{experience replay}~\cite{robins1995,rolnick2019,chaudhry2019efficient}. The fix is mechanical: mix old-domain data into the new-domain training loss, constraining the optimizer not to destroy the old skill. The rest of the paper quantifies how much such mixing is needed.

The paper is organized as follows. Section~\ref{sec:bg} lays out the necessary background. Section~\ref{sec:related} reviews related work. Section~\ref{sec:data} describes the data. Section~\ref{sec:method} specifies the architecture and training. Section~\ref{sec:results} reports results. Section~\ref{sec:discussion} discusses implications. Section~\ref{sec:limitations} lists limitations. Section~\ref{sec:conclusion} concludes.

\section{Background}
\label{sec:bg}

\subsection{Chord progression generation as a sequence task}
\label{sec:bg-task}

A chord progression is a finite sequence of chord labels indexed against a musical timeline. Each label is structured: a root, a quality, optional extensions, and an optional bass for slash chords like \texttt{C/G}. Modern systems flatten this into a vocabulary of \emph{chord tokens}, one (root, quality) pair per token, and treat sequence modeling over that vocabulary as the learning problem~\cite{makris2020chord,dalmazzo2024chordinator,paiement2005probabilistic}. Palette size varies. Triads-and-sevenths palettes of 24 to 48 tokens are common in Bach-chorale-style work~\cite{hadjeres2017deepbach,liang2017bachbot}. Palettes for extended jazz qualities (\texttt{maj9}, \texttt{m11}, \texttt{13\#11}) reach several hundred.

This paper uses 351 tokens: 312 chord tokens (twelve roots $\times$ twenty-six qualities), twenty-four key signatures (twelve major, twelve minor), five time-signature markers, six genre markers, and four structural tokens (BOS, EOS, BAR, padding). It covers all 52.2M chord events in the union of the six datasets after a normalization pass that drops slash-chord basses, reconciles alternate quality notations (\texttt{Cmin7} vs.\ \texttt{C:min7}) and JAAH's interval-based notation, and maps the spellings no alias rule covers (0.8\% of chord events) with string heuristics (Section~\ref{sec:data-tokenizer}).

Within this representation, chord generation is standard autoregressive sequence modeling. The Music Transformer~\cite{huang2019music} was designed for polyphonic note-event prediction, but applying it to chord events is a smaller problem: smaller per-step vocabulary, shorter sequences (tens to a few hundred chord events per song versus thousands of note events), no polyphony. Relative-position attention remains useful because chord progressions have regular periodic structure (8- and 16-bar phrases, AABA forms, sectional repetition).

Two evaluation issues recur. First, top-$k$ next-token accuracy on held-out chord sequences does not measure musicality directly. It measures alignment with human transcribers. A model that always predicts the most common diatonic substitute will outperform one that ventures interesting modal interchange, because the data over-represents diatonic continuations. I treat top-1 and top-5 as a \emph{proxy} for whether the model has learned a genre's local statistics. The proper complement is a controlled listening study with multiple raters, which this paper does not include and which Sections~\ref{sec:limitations} and~\ref{sec:conclusion} identify as the natural next experiment. Second, jazz and pop test sets have different per-token entropy. Jazz harmonic vocabulary spans more tokens with flatter usage, so the \emph{baseline} jazz accuracy of a strong pop-pretrained model is lower (72.86\%) than baseline pop accuracy (84.21\%). I reason in terms of \emph{changes} relative to the per-genre baseline.

\subsection{Pop and jazz harmonic vocabulary}
\label{sec:bg-vocab}

The pop and jazz training corpora share a large core: major and minor triads on diatonic degrees, dominant sevenths, and sus chords. They diverge in three regions that turn out to be the active difference between the genres in the experiments below. Modal interchange (e.g.\ \texttt{iv} borrowed from the parallel minor) is occasional in pop and routine in jazz~\cite{declercq2011corpus}. Secondary dominants and tritone substitutions (\texttt{bII7} for \texttt{V7}) are jazz devices that rarely appear in pop transcriptions. Long II--V chains pivoting through several keys are similar. Extended seventh chords (\texttt{maj9}, \texttt{13}, \texttt{m11}) appear as chord-symbol notation in jazz lead sheets but get simplified to seventh-chord skeletons in pop transcriptions, even when the recording's voicing has the upper extensions.

Pop and jazz are therefore not disjoint vocabularies. Most chord \emph{tokens} are shared. They have meaningfully different \emph{transition statistics} over those tokens. A pop-trained model emits short cycles of diatonic chords with occasional modal mixture. A jazz-trained model favors descending fifth motion (II--V--I), substitutional chains, and extensions. Forgetting in the pop-to-jazz direction is not the loss of any particular token but a gradual reweighting of $P(\text{next} \mid \text{context})$ toward jazz-typical continuations. Recovery requires preserving the original transition statistics, which is what rehearsal does.

Because the genres share most tokens, the per-genre top-1 gap is bounded. A strong pop model (Phase~0, Section~\ref{sec:method-phase0}) already gets approximately 73\% of jazz test targets right. Bar tokens supply 39 of those points, since it predicts them 95\% of the time, while on the jazz chord targets alone it is right 60\% of the time, mostly on the diatonic continuations both genres share, and chord targets account for 22 of the 27 points it misses. Any successful jazz fine-tune should lift jazz accuracy 5 to 10 points, and the runs here do. The interesting axis is what happens to \emph{pop} accuracy as the model acquires that jazz vocabulary. That is what this sweep measures.

\subsection{Catastrophic forgetting and rehearsal}
\label{sec:bg-cf}

Catastrophic forgetting was first documented by~\cite{mccloskey1989catastrophic} and reviewed in~\cite{french1999catastrophic,goodfellow2013empirical}. A network trained sequentially on tasks A and B degrades sharply on A after training on B. The phenomenon is severe when the two loss surfaces share parameter regions in conflicting ways and mild when both tasks admit simultaneous solutions in the same region. Pop and jazz chord generation, sharing tokens but differing in transition statistics, sit in the middle. There is no architectural reason both distributions cannot live in one model, but SGD updates on jazz alone do drift the parameters into pop-degraded regions, as the F5 (jazz-only) results show.

Continual learning~\cite{de2021continual,wang2024comprehensive} distinguishes three families. \emph{Regularization} approaches penalize updates that damage the prior task's loss surface. Elastic Weight Consolidation~\cite{kirkpatrick2017overcoming} weights updates by Fisher information for the prior task. Learning without Forgetting~\cite{li2018learning} adds a knowledge-distillation term against a frozen teacher. \emph{Parameter isolation} approaches allocate disjoint parameter subsets per task, including progressive networks and adapter-based fine-tuning. \emph{Rehearsal} or \emph{experience replay} approaches~\cite{robins1995,rebuffi2017icarl,rolnick2019,chaudhry2019efficient} mix old-task samples into the new-task loss. Some formulations constrain the rehearsal buffer to satisfy a per-task gradient projection criterion~\cite{lopez2017gradient}.

I use rehearsal. It has a single knob, the rehearsal buffer size, which makes the question (\emph{how much rehearsal is enough}) directly answerable by varying that quantity. The mixed pop-jazz training data is already available, so the engineering cost is essentially zero. Rehearsal also produces a single model that fluently generates both genres at inference, which is what the application needs. Parameter-isolation routes would complicate model serving without a clear performance benefit at this scale.

A related literature is \emph{data mixture} in language modeling. The Pile~\cite{gao2020pile} used a 22-corpus heuristic mixture, and DoReMi~\cite{xie2023doremi} introduced learned proxy methods for choosing better mixtures. These works concern \emph{pretraining}, where the goal is one foundation model across many domains. The work here operates at \emph{fine-tuning} with a fixed pretrain. The asymmetry in this setting (the pop corpus is $366\times$ the jazz corpus) also differs from the symmetric-availability assumption these works typically make. The underlying claim that proportions are a meaningful empirical lever still applies.

\section{Related Work}
\label{sec:related}

\subsection{Symbolic chord-progression models}
\label{sec:related-models}

Chord-progression generation predates deep learning. Early statistical work used Markov chains and probabilistic context-free grammars~\cite{steedman1984generative,rohrmeier2011generative,paiement2005probabilistic}. These captured local dependence and (in the grammar case) hierarchy but not the long-range thematic recurrence of commercial songwriting. Granroth-Wilding and Steedman~\cite{granroth2014robust} introduced an early annotated jazz chord corpus and a parser-interpreter, since substantially superseded by the Jazz Harmony Treebank~\cite{harasim2020jht}.

Recurrent and convolutional architectures replaced grammars in the late 2010s, usually as components of full-song pipelines~\cite{yang2017midinet,roberts2018hierarchical,dong2018musegan} rather than standalone chord generators. ChordRipple~\cite{huang2016chordripple} framed chord substitution as co-creative interaction, suggesting alternatives rather than autocompleting. That framing influences the application's UI.

Three transformer-based systems are direct comparators. Chord Jazzification~\cite{makris2020chord} learns jazz voicings conditioned on chord symbols (voicing-level, not progression-level). BachBot~\cite{liang2017bachbot} and DeepBach~\cite{hadjeres2017deepbach} target chorale harmonization, but the underlying autoregressive formulation is the same. The Chordinator~\cite{dalmazzo2024chordinator} is closest in spirit: a transformer trained on multi-genre chord sequences with style conditioning. It demonstrates style-conditioned generation across genres but does not study the per-genre transfer trade-off under fine-tuning, and its evaluation aggregates accuracy across genres. This work explicitly holds the target genre fixed and sweeps the source-genre rehearsal volume, with per-genre evaluation.

\subsection{Genre adaptation and transfer in symbolic music}
\label{sec:related-transfer}

Cross-genre transfer is well-attested in audio MIR~\cite{choi2017transfer}. Symbolic music has fewer published comparisons because symbolic corpora are smaller and the question of which corpora are large enough to serve as a useful source domain is open. Hung et al.~\cite{hung2019improving} report pop-to-jazz melody transfer with a variational RNN. Score Transformer~\cite{suzuki2021score} demonstrates pop-to-classical fine-tuning for score prediction. Both use a single pure-target fine-tune step, which corresponds to F5 (jazz-only) in the experiments here. F5 is the worst configuration on the source-genre axis without offering target-genre benefit.

The Continuator~\cite{pachet2017continuator} and similar interactive style-imitation systems represent an older non-deep-learning thread. They adapt to a single user's input style rather than transferring between large-scale corpora, but they share the practical concern that \emph{genre fluency} is the behavioral target rather than any single accuracy metric.

\subsection{Continual learning, rehearsal, and forgetting}
\label{sec:related-cf}

The conceptual frame comes from continual learning. Beyond the foundational works named in Section~\ref{sec:bg-cf}, this setting differs from the typical continual-learning benchmark in two ways. I do not impose a gradient-projection constraint~\cite{lopez2017gradient,chaudhry2019efficient}. I just mix old-domain data into the new-domain loss in a fixed proportion. And I do not vary the architecture across tasks (no progressive columns, no adapters). The empirical question I address (sample-complexity in this simple-mixture regime) has not, to my knowledge, been quantified for symbolic chord generation. Curriculum learning~\cite{bengio2009curriculum} is a related but distinct idea that orders examples by difficulty rather than mixing across domains. I mix. I do not order.

\subsection{Data mixture in language model pretraining}
\label{sec:related-mix}

The proportion question has received serious attention in language modeling. The Pile~\cite{gao2020pile} documents a human-chosen 22-corpus mixture that became a \emph{de facto} standard for early open LLMs. DoReMi~\cite{xie2023doremi} introduces learned proxy-model mixture selection. Both operate at pretraining, where the goal is one foundation model across many tasks. The work here operates at fine-tuning with a fixed pretrain. The asymmetry in this setting (pop corpus $366\times$ jazz corpus) also differs from the symmetric-availability assumption these works typically make. The underlying claim that proportions are a meaningful lever still applies.

\subsection{Position of the present work}
\label{sec:related-pos}

Table~\ref{tab:positioning} summarizes the position of this work relative to the prior literature on related tasks.

\begin{table}[h]
\centering
\caption{Positioning of this work relative to prior literature.}
\label{tab:positioning}
\resizebox{\linewidth}{!}{%
\begin{tabular}{lllll}
\toprule
Work & Primary task & Genre adaptation & Mix sweep & Per-genre eval \\
\midrule
Granroth-Wilding \& Steedman 2014 & jazz chord parsing & --- & --- & --- \\
Hung et al.\ 2019 & melody generation & pop $\to$ jazz & no & melody metrics \\
Suzuki 2021 (Score Transformer) & score prediction & pop $\to$ classical & no & token-level \\
Dalmazzo et al.\ 2024 (Chordinator) & chord progression & multi-genre joint & no & aggregate \\
Hadjeres et al.\ 2017 (DeepBach) & chorale harmonization & none & --- & corpus-internal \\
\textbf{This work} & \textbf{chord progression} & \textbf{pop $\to$ jazz} & \textbf{5 mix sizes} & \textbf{pop + jazz split} \\
\bottomrule
\end{tabular}}
\end{table}

The contribution is the specific intersection of (i) chord-progression generation as the primary task, (ii) cross-genre fine-tuning, (iii) a systematic rehearsal-mix sweep including the all-or-nothing endpoints, and (iv) per-genre held-out evaluation that lets the trade-off be read directly from a single table. Each design choice individually has prior art. The contribution is the combination and the bound I report on how much rehearsal is needed.

\section{Data}
\label{sec:data}

\subsection{Source corpora}
\label{sec:data-corpora}

The six corpora were chosen for the largest pop and jazz chord-symbol datasets available under research-permitting licenses, with minimal cross-corpus overlap. Pop comes from two sources. \emph{Chordonomicon}~\cite{kantarelis2024chordonomicon} is a user-generated corpus of approximately 679{,}000 chord-annotated songs covering pop, rock, and adjacent commercial genres. I use it as the primary pop corpus. McGill \emph{Billboard}~\cite{burgoyne2011} adds approximately 890 expert-annotated transcriptions of charted pop songs as a high-quality reference subset. The pop pool is used essentially as published: light filtering (rows with fewer than four usable chord tokens) removes only approximately 0.05\% of Chordonomicon, leaving a 680{,}373-song pop pool. Chordonomicon's main-genre tags are mapped onto three markers: rock, alternative, punk, and metal give \texttt{[GENRE:rock]}, jazz gives \texttt{[GENRE:jazz]}, and every other tag (pop, pop rock, country, soul, rap, reggae, electronic) or a missing one, as for nearly half of the songs, gives \texttt{[GENRE:pop]}, as do all Billboard songs. Of the 544{,}272 pop-pool training songs, 425{,}274 therefore carry \texttt{[GENRE:pop]} (715 of them from Billboard), 113{,}389 carry \texttt{[GENRE:rock]}, and 5{,}609 carry \texttt{[GENRE:jazz]}. Chordonomicon has no key annotation, so its songs carry the placeholder key \texttt{[KEY:Cmaj]} (Section~\ref{sec:method-eval}).

Jazz comes from four sources. The \emph{Jazz Harmony Treebank} (JHT)~\cite{harasim2020jht} contributes approximately 1{,}170 expertly annotated jazz standards, derived from iReal Pro with hand correction. \emph{JazzStandards}~\cite{jazzstandards_irealpro} adds approximately 293 standards after JHT-deduplication. \emph{Weimar Jazz Database} (WJazzD)~\cite{pfleiderer2017jazzomat} adds approximately 283 chord-annotated jazz solos. \emph{JAAH}~\cite{eremenko2018jaah} adds 113 audio-aligned transcriptions from the Smithsonian collection. Only JazzStandards is deduplicated against JHT (by title, since both derive from iReal Pro), and WJazzD contributes one chord sequence per composition, so the same tune can appear in more than one source. The total jazz corpus is 1{,}859 songs, of which 1{,}513 are in the train split.

\begin{table}[h]
\centering
\small
\caption{Training data sources and song counts as used (JazzStandards deduplicated against JHT, pop lightly filtered).}
\label{tab:data}
\begin{tabular}{llrl}
\toprule
Genre & Dataset & Songs & License \\
\midrule
\multirow{2}{*}{Pop}
  & Chordonomicon~\cite{kantarelis2024chordonomicon} & 679{,}483 & CC BY-NC 4.0 \\
  & McGill Billboard~\cite{burgoyne2011} & 890 & CC0 \\
\midrule
\multirow{4}{*}{Jazz}
  & Jazz Harmony Treebank~\cite{harasim2020jht} & 1{,}170 & Public \\
  & JazzStandards (iReal Pro)~\cite{jazzstandards_irealpro} & 293 & Community \\
  & Weimar Jazz Database~\cite{pfleiderer2017jazzomat} & 283 & ODbL \\
  & JAAH~\cite{eremenko2018jaah} & 113 & Research \\
\midrule
\multicolumn{2}{l}{Total jazz} & 1{,}859 & \\
\bottomrule
\end{tabular}
\end{table}

The corpus-size asymmetry (approximately 680K pop versus approximately 1.9K jazz) is itself a structural feature of the setup. Pop and rock have been transcribed by hobbyists at scale, while jazz has been curated by smaller research and pedagogical communities. Any practical genre-adaptation pipeline targeting jazz from a pop pretrain faces this asymmetry, and the sweep here is designed around it. Because the jazz pool is small enough to be exhausted, every fine-tune run uses the entire jazz training split. The variable is the volume of pop rehearsal mixed alongside, with sizes (1K, 2.5K, 5K, 10K) spanning from less than the jazz training volume up to roughly seven times it.

\subsection{Notation harmonization and the unified tokenizer}
\label{sec:data-tokenizer}

Chord notation is inconsistent across the six corpora. Chordonomicon uses guitar-oriented \texttt{Cmaj7}, \texttt{Cmin7}, \texttt{C7}, with \texttt{s} for a sharp (\texttt{Fs7}). JHT uses iReal Pro's \texttt{C\^{}7}, \texttt{Cm7}, \texttt{C\%7}, and JazzStandards spells \texttt{Cmaj7}, \texttt{Cm7}, \texttt{C7} in plain text. WJazzD uses a SQLite-encoded chord class field. JAAH uses interval-based notation. Slash chords (\texttt{C/G}), enharmonic variants (\texttt{Db} vs.\ \texttt{C\#}), and rare qualities (e.g.\ \texttt{mMaj7}, \texttt{add\#11}) appear inconsistently. The chord-normalization pass maps every input to a canonical (root, quality) tuple from a 12-root, 26-quality palette, giving 312 chord tokens. Adding twenty-four key-signature tokens (twelve major, twelve minor), five time-signature tokens, six genre markers (\texttt{[GENRE:jazz]}, \texttt{[GENRE:pop]}, \texttt{[GENRE:rock]}, \texttt{[GENRE:blues]}, \texttt{[GENRE:bossa]}, \texttt{[GENRE:none]}), and four structural tokens (\texttt{BOS}, \texttt{EOS}, \texttt{BAR}, \texttt{PAD}) gives the final 351-token vocabulary ($312 + 24 + 5 + 6 + 4$).

Every one of the 52.2M chord events maps to a palette token after normalization. Spellings that no alias rule covers (0.8\% of chord events) go through string heuristics that do not always keep the chord family: a spelling that starts with \texttt{7} and carries a b9, \#9, \#11, or b13 becomes that altered dominant, any other spelling that starts with \texttt{m} or a minus sign becomes a minor token (so \texttt{maj13} becomes \texttt{m13}), and everything else becomes the major triad (\texttt{7sus4} and \texttt{13b9} included). About 0.2\% of all chord events thereby leave their spelled family, mostly \texttt{7sus4} chords that become major triads, and JAAH interval sets outside its conversion table default to the dominant seventh. Two normalization choices: enharmonically equivalent roots are collapsed (\texttt{Db} and \texttt{C\#} map to one token, consistent with iReal Pro practice), and slash-chord basses are dropped (\texttt{C/G} becomes \texttt{C}), so inversions are not represented. Both reduce vocabulary size and, in inspection of generated outputs, did not surface audible cases where the collapse mattered.

\subsection{Splits and augmentation}
\label{sec:data-splits}

The pooled song list of all six corpora is split 80/10/10 at the \emph{song} level (not the chord-event level) in one shuffle with seed 42, so no song record crosses the boundary, although a tune held by two jazz sources can fall on both sides (29 of the 167 jazz test songs share their title with a jazz training song from another source), and 6{,}060 of the 68{,}057 pop test songs (8.9\%) have the same chord sequence as a pop training song. The shuffle is not stratified by corpus, so per-corpus proportions vary slightly: the 1{,}859 jazz songs fall into 1{,}513 train, 179 validation, and 167 test songs. Training sequences are augmented by twelve-key transposition. Validation and test sequences are not transposed. Held-out test sets are filtered by source corpus so that pop test (Chordonomicon + Billboard) and jazz test (JHT + WJazzD + JAAH + JazzStandards) can be evaluated independently. All Section~\ref{sec:results} numbers are on these per-genre held-out sets.

\section{Method}
\label{sec:method}

\subsection{Architecture}
\label{sec:method-arch}

Music Transformer~\cite{huang2019music} with relative-position attention. $d_{\text{model}} = 512$, eight heads, $d_{\text{ff}} = 2048$, eight layers, max sequence length 256 (longer songs are cut to their first 255 tokens followed by the end token), dropout 0.1. Total parameters: 25{,}661{,}440. The size is small relative to recent symbolic music transformers like Compound Word Transformer~\cite{hsiao2021compound} and Anticipatory Music Transformer~\cite{thickstun2024anticipatory}, but well-matched to chord-only sequence modeling (vocabulary two orders of magnitude smaller than full polyphonic note-events) and to the consumer GPU budget I trained on (one NVIDIA RTX 4070 Mobile, 8 GB VRAM).

\subsection{Phase 0: pop pretraining}
\label{sec:method-phase0}

Train from scratch on the pop training split (Chordonomicon + Billboard: the approximately 544K-song train split, 80\% of the 680{,}373-song pop pool; approximately 6.5M sequences after twelve-key augmentation). Three epochs, micro-batch 64 with gradient accumulation to effective batch 128, AdamW, peak lr $3 \times 10^{-4}$ under a per-epoch cosine schedule (epochs 0 and 1 at $3 \times 10^{-4}$, epoch 2 at $1.5 \times 10^{-4}$, since the configured one-epoch warmup has no effect when the rate is set per epoch), fp16. Wall-clock approximately 27 hours on the RTX 4070 Mobile. Best-epoch metrics on held-out pop test: 84.21\% top-1, 97.09\% top-5. Same checkpoint on jazz test: 72.86\% top-1, 86.51\% top-5, reflecting the substantial token overlap.

The Phase 0 checkpoint is the starting point for all five fine-tune runs.

\subsection{Phase 1: jazz fine-tuning with pop rehearsal}
\label{sec:method-phase1}

Five fine-tune experiments F1 through F5, each resuming from the Phase 0 best checkpoint. Each trains on \emph{all} 1{,}513 jazz training sequences plus a varying number of pop rehearsal sequences sub-sampled with a fixed seed. Table~\ref{tab:runs} lists the configurations.

\begin{table}[h]
\centering
\small
\caption{Fine-tune experiment configurations. Jazz volume is fixed at all 1{,}513 training songs; pop mix is swept.}
\label{tab:runs}
\begin{tabular}{llrrrr}
\toprule
Run & Config & Pop mix & Jazz & Total & Pop fraction \\
\midrule
F1 & \texttt{ft\_jazz\_pop80} & 10{,}000 & 1{,}513 & 11{,}513 & 87\% \\
F2 & \texttt{ft\_jazz\_pop67} &  5{,}000 & 1{,}513 &  6{,}513 & 77\% \\
F3 & \texttt{ft\_jazz\_pop50} &  2{,}500 & 1{,}513 &  4{,}013 & 62\% \\
F4 & \texttt{ft\_jazz\_pop29} &  1{,}000 & 1{,}513 &  2{,}513 & 40\% \\
F5 & \texttt{ft\_jazz\_only}  &      0   & 1{,}513 &  1{,}513 &  0\% \\
\bottomrule
\end{tabular}
\end{table}

Common hyperparameters as configured: ten epochs maximum with early stopping (patience 5), peak lr $2 \times 10^{-5}$, two-epoch warmup, otherwise identical to Phase 0. As applied, every run resumed with the Phase~0 optimizer and scheduler state, which overrode the configured rate and warmup. The restored learning rate was 0 because the Phase~0 cosine had fully decayed, so the first fine-tune epoch (epoch~3) updated no weights and its metrics equal the Phase~0 endpoint. From epoch~4 the fine-tune's cosine schedule ran on the restored base rate of $3 \times 10^{-4}$: $2.56 \times 10^{-4}$, $2.07 \times 10^{-4}$, $1.50 \times 10^{-4}$, $9.26 \times 10^{-5}$, $4.39 \times 10^{-5}$, and $1.14 \times 10^{-5}$ for epochs 4 to 9. F2, F3, and F4 ran to epoch~9, F1 stopped after epoch~8 by early stopping, and the F5 log ends at epoch~7. All five runs share this applied schedule, so the sweep compares rehearsal volumes under one schedule, but it is not the configured one. Twelve-key augmentation applies during fine-tune as in pretrain.

\subsection{Evaluation protocol}
\label{sec:method-eval}

After each epoch the checkpoint is evaluated on three held-out splits: the validation split of the fine-tune mix, the full pop test set, and the full jazz test set. I record cross-entropy loss, perplexity, top-1, and top-5 per split. All four are micro-averages over every non-padding target token, so bar markers, the end token, and the three header tokens that open every sequence (key, time signature, and genre marker) count alongside chords. Chord tokens are 64\% of the pop and 55\% of the jazz test targets, and each header token is 0.86\% and 1.07\%. A checkpoint predicts each header token from the header tokens before it, so it makes nearly the same header prediction for every test sequence, and each header target moves whole runs at once. Every checkpoint predicts the 4/4 time signature. Phase~0 and the pop-majority runs F1 to F3 predict \texttt{[GENRE:pop]} for nearly every sequence, while the jazz-majority runs F4 and F5 predict \texttt{[GENRE:jazz]} for every sequence (measured on the released checkpoints, and steady from epoch~4 on, since their pop curves show no step of the size this marker is worth), which costs them 0.67 pop points and gains them 1.07 jazz points against Phase~0. Chordonomicon has no key annotation, so every Chordonomicon song carries the placeholder \texttt{[KEY:Cmaj]}, as do the WJazzD and JazzStandards songs. JHT writes a flat as a minus sign (\texttt{E-}, \texttt{B-}), which the key parser does not read, so its 445 songs in flat keys fall back to \texttt{[KEY:Cmaj]} as well, and its minor keys become the major key on the same tonic (one JAAH test song with an unread key spelling also falls back). \texttt{[KEY:Cmaj]} is thus the key target of 67{,}983 of the 68{,}057 pop test sequences, 6 of them Billboard songs annotated in C major, and of 109 of the 167 jazz test sequences, 16 of them annotated in C major. Twelve-key augmentation makes the key prediction after \texttt{BOS} nearly uniform (Phase~0 gives B$\flat$ major 0.093 and C major 0.087), so which major key a checkpoint predicts there is close to arbitrary, and an epoch that predicts C major scores about 0.86 more pop points and 0.70 more jazz points. None of the released checkpoints does. Three seed retrains of the F1 recipe that kept every epoch's checkpoint (on the expanded corpus of the v2 Correction) predict C major at exactly the four of their 18 updating epochs whose pop and jazz top-1 jump together, and the same jump marks the tabulated epochs of F1 (epoch~6) and F3 (epoch~9), whose pop top-1 sits about 0.8 points and jazz top-1 0.6 to 0.8 points above the neighboring epochs. Table~\ref{tab:headerfree} therefore also gives top-1 without the three header targets, over bar, chord, and end targets only, on which Phase~0 scores 84.87 pop and 74.22 jazz. Only the selected checkpoint of each run was kept, so the table removes the header targets from the logged all-target top-1, with header counts measured on the released checkpoints and, at the other epochs, taken from the marker each run predicts and from the C-major jumps. Rescoring the released checkpoints on the jazz test set reproduces the derived jazz values within 0.02 points. All numbers are written per epoch to \texttt{eval\_results.csv}.

For each run I report metrics at the \emph{best-performing epoch}: the epoch with the highest jazz top-1 subject to pop top-1 staying within 3 points of the Phase 0 baseline. The constraint reflects the goal of acquiring jazz capability \emph{without} sacrificing pop fluency. Without it, the constraint-free best-epoch would track jazz top-1 monotonically and mask the catastrophic forgetting I want to characterize. In practice the constraint is non-binding for all five runs: the lowest pop top-1 at any epoch (82.1, for F5) stays above the threshold. For F2 and F5, however, the tabulated row is the released minimum-mixed-validation-loss checkpoint (see Correction), which differs from the jazz-top-1 maximum by at most half a point and does not affect the sweep conclusions. Because the key target moves jazz top-1 by 0.70 points, the rule picks an epoch that predicts C major wherever a run has one, as F1 and F3 do.

Each released checkpoint (the minimum mixed-validation-loss selection described in the Correction; for the jazz-adapted F1 endpoint, the selection-corrected retrain \texttt{ft-pop80-v2}) also generates three sample continuations under top-$p = 0.9$ and temperature 0.8 for five prompts: pop I--vi--ii--V in C, pop I--V--vi--IV in G, jazz ii--V--I in C, jazz turnaround in B$\flat$, and jazz minor ii--V resolving to A minor. These continuations were inspected by the author. \textbf{No formal listening study with multiple raters was conducted in this version of the paper.}

\section{Results}
\label{sec:results}

\subsection{Headline metrics}
\label{sec:results-headline}

\begin{figure}[h]
  \centering
  \includegraphics[width=0.85\linewidth]{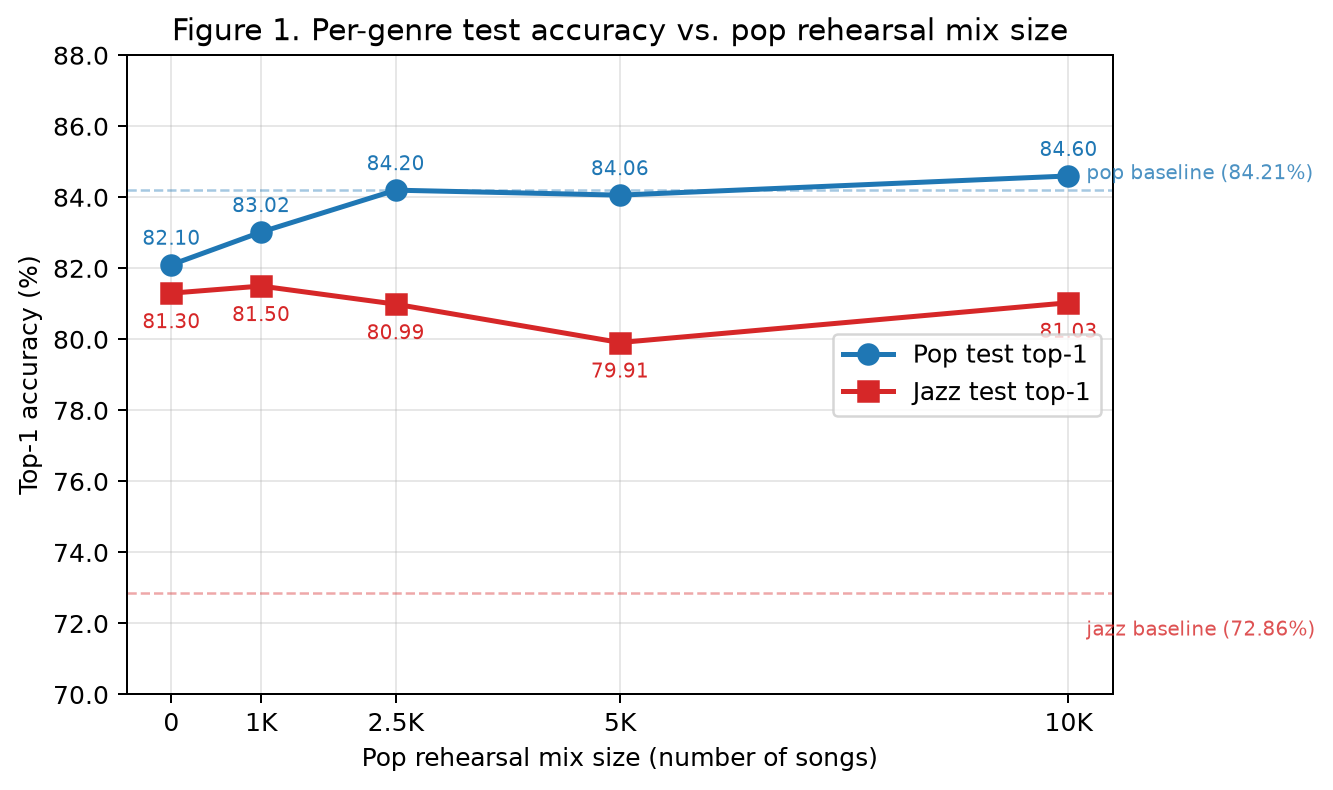}
  \caption{Per-genre top-1 next-token accuracy over all targets at each run's tabulated epoch, by pop rehearsal mix size. Dashed lines mark the Phase 0 baseline. The F1 and F3 points are epochs that predict the C-major key, which adds about 0.86 pop and 0.70 jazz points, and the F4 and F5 points include a 0.67-point pop loss and a 1.07-point jazz gain from predicting the jazz marker (Section~\ref{sec:method-eval}). Table~\ref{tab:headerfree} gives the values without the header targets.}
  \label{fig:1}
\end{figure}

Table~\ref{tab:main} reports best-epoch per-genre test accuracy for each run. Table~\ref{tab:deltas} reports the same numbers as deltas relative to the Phase 0 baseline. Figure~\ref{fig:1} visualizes the per-genre accuracy as a function of pop rehearsal volume, and Table~\ref{tab:headerfree} gives the deltas without the three header targets, by epoch.

\begin{table}[h]
\centering
\small
\caption{Per-genre test metrics over all targets at each run's best epoch. The top-5 columns carry a key effect of the same kind (Section~\ref{sec:method-eval}): pop top-5 counts the key target only when C major is among a checkpoint's five most likely keys after \texttt{BOS}, and the F2 checkpoint, whose C major ranks eleventh, has the lowest pop top-5 in the table.}
\label{tab:main}
\begin{tabular}{lrrrrrr}
\toprule
Run & Pop mix & Pop top-1 & Pop top-5 & Jazz top-1 & Jazz top-5 & Jazz ppl \\
\midrule
Phase 0 (ep 2) & --- & \textbf{84.21} & 97.09 & 72.86 & 86.51 & 4.01 \\
F1 (ep 6) & 10{,}000 & \textbf{84.60} & 96.96 & 81.03 & 92.41 & 2.31 \\
F2 (ep 4) &  5{,}000 & 84.06 & 96.17 & 79.91 & 91.46 & 2.34 \\
F3 (ep 9) &  2{,}500 & 84.20 & 96.87 & 80.99 & 92.63 & 2.29 \\
F4 (ep 6) &  1{,}000 & 83.02 & 96.93 & \textbf{81.50} & 92.59 & 2.26 \\
F5 (ep 7) &      0   & 82.10 & 96.31 & 81.30 & 92.44 & 2.24 \\
\bottomrule
\end{tabular}
\end{table}

\begin{table}[h]
\centering
\small
\caption{Deltas relative to Phase 0 baseline over all targets (positive = better). Table~\ref{tab:headerfree} gives them without the header targets.}
\label{tab:deltas}
\begin{tabular}{lrr}
\toprule
Run & $\Delta$ Pop top-1 & $\Delta$ Jazz top-1 \\
\midrule
F1 (10K mix) & \textbf{+0.39} & +8.17 \\
F2 (5K mix)  & $-$0.15 & +7.05 \\
F3 (2.5K mix)& $-$0.01 & +8.13 \\
F4 (1K mix)  & $-$1.19 & \textbf{+8.64} \\
F5 (jazz only)& $-$2.11 & +8.44 \\
\bottomrule
\end{tabular}
\end{table}

\begin{table}[h]
\centering
\small
\caption{Top-1 deltas relative to Phase 0 without the three header targets (bar, chord, and end targets only, Section~\ref{sec:method-eval}): pop at every updating epoch, and pop and jazz at each run's tabulated epoch. F4's jazz value is a range because the key its epoch-6 checkpoint predicts is not known.}
\label{tab:headerfree}
\begin{tabular}{lrrrrrrrr}
\toprule
 & \multicolumn{6}{c}{$\Delta$ Pop top-1 by epoch} & \multicolumn{2}{c}{Tabulated epoch} \\
Run & 4 & 5 & 6 & 7 & 8 & 9 & $\Delta$ Pop & $\Delta$ Jazz \\
\midrule
F1 (10K mix) & $-$0.14 & $-$0.30 & $-$0.49 & $-$0.64 & $-$0.73 & --- & $-$0.49 & +7.67 \\
F2 (5K mix) & $-$0.15 & $-$0.35 & $-$0.53 & $-$0.69 & $-$0.82 & $-$0.84 & $-$0.15 & +7.19 \\
F3 (2.5K mix) & $-$0.15 & $-$0.35 & $-$0.56 & $-$0.75 & $-$0.84 & $-$0.90 & $-$0.90 & +7.63 \\
F4 (1K mix) & $-$0.17 & $-$0.34 & $-$0.54 & $-$0.69 & $-$0.83 & $-$0.88 & $-$0.54 & +7.6 to +7.8 \\
F5 (jazz only) & $-$0.72 & $-$1.07 & $-$1.36 & $-$1.48 & --- & --- & $-$1.48 & +7.62 \\
\bottomrule
\end{tabular}
\end{table}

Four observations follow.

First, every fine-tuned run acquires jazz capability at its best epoch. Jazz top-1 jumps from 72.86\% baseline to a band of 79.91 to 81.50\% across the five runs, a 7- to 9-point gain. These are best-epoch metrics: the released checkpoints, selected by minimum mixed-validation loss, differ for F1/F3/F4, and the released F1 coincides with Phase~0 (see Correction). Variation within that band is small (approximately 1.6 points between F2 and F4) and does not correlate monotonically with rehearsal volume. Without the header targets the gain is 7.2 to 7.8 points (Table~\ref{tab:headerfree}), F2's lower value reflecting its earlier epoch, at which the other rehearsal runs gain 7.0 to 7.2. Whatever else fine-tuning does, it consistently transfers jazz harmonic statistics.

Second, on the logged metric pop is preserved across most of the sweep but collapses at the no-rehearsal end: F5 drops 2.11 points, F4 1.19, F2 0.15, and F3 0.01, and F1 gains 0.39 at its best epoch. Most of that spread comes from the header targets (Section~\ref{sec:method-eval}). F4 and F5 lose 0.67 points to the jazz marker, and the tabulated epochs of F1 and F3 gain 0.86 from the C-major key. Without the header targets F5 drops 1.48 points, while the four rehearsal runs drop 0.15 to 0.90 points depending mainly on their tabulated epoch, and at any one epoch they lie within 0.11 points of each other, the 10K run losing the least (Table~\ref{tab:headerfree}). The protective step is therefore between no rehearsal and the smallest volume tested, 1K rehearsal sequences or $0.66\times$ the jazz training volume. The released F1 in any case coincides with Phase~0 and gains nothing (see Correction).

Third, returns saturate from the smallest volume tested. Without the header targets, raising rehearsal from 1K to 10K reduces the pop loss by at most 0.11 points at any epoch and leaves the jazz gain within the same band. The 10K run loses the least pop at every epoch, while 1K, 2.5K, and 5K show no order by volume, so on this metric no volume above 1K buys much for its extra training cost.

Fourth, jazz-only fine-tuning buys no jazz advantage. Without the header targets its jazz gain ($+7.62$) lies inside the band of the rehearsal runs, while its pop loss at every epoch is about twice theirs or more. F5 is a calibration of the no-rehearsal failure mode, not a viable production checkpoint.

\subsection{Learning dynamics}
\label{sec:results-dynamics}

\begin{figure}[h]
  \centering
  \includegraphics[width=\linewidth]{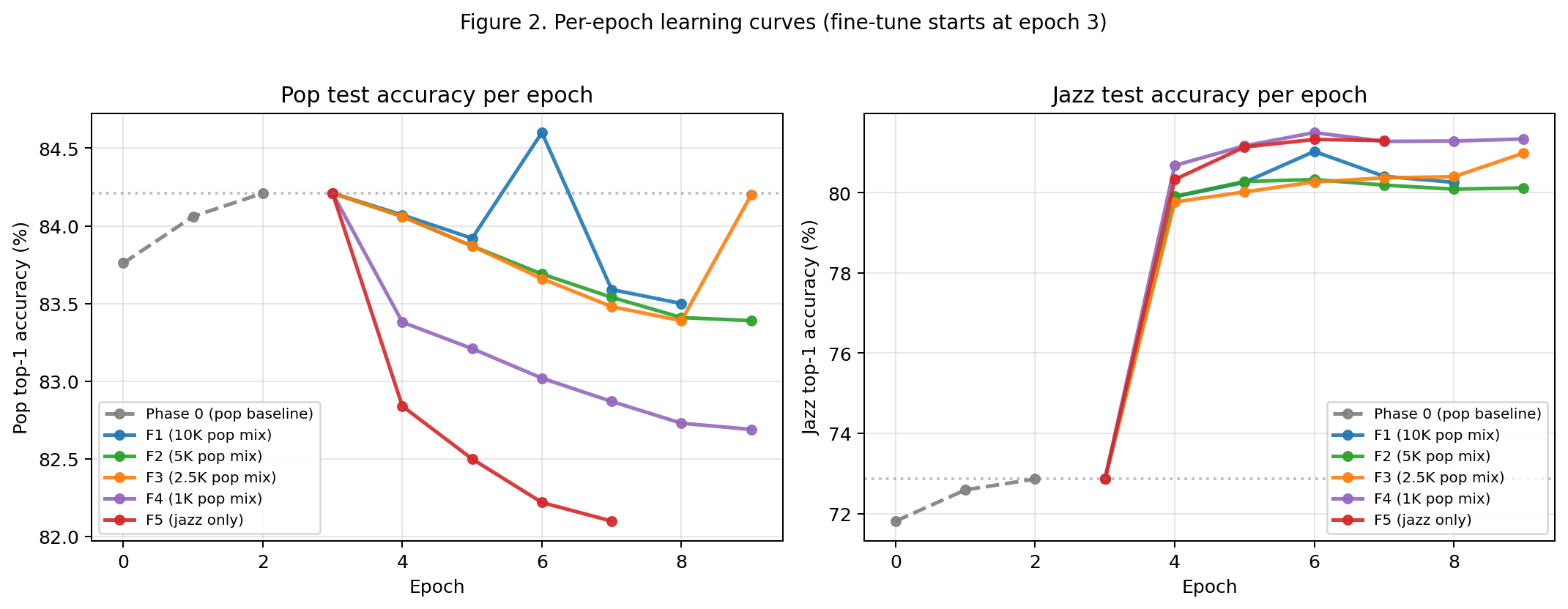}
  \caption{Per-epoch pop (left) and jazz (right) top-1 accuracy across all runs. Phase 0 is dashed gray, and fine-tune runs branch from epoch 3. Single-epoch rises of about 0.8 pop and 0.7 jazz points (F1 at epoch 6, F3 at epoch 9) are epochs that predict the C-major key, and the F4 and F5 curves include a 0.67-point pop loss and a 1.07-point jazz gain from predicting the jazz marker (Section~\ref{sec:method-eval}).}
  \label{fig:2}
\end{figure}

Figure~\ref{fig:2} plots per-epoch top-1 for pop and jazz across all six runs. Phase 0 occupies epochs 0 to 2 as a dashed gray reference. The five fine-tunes branch from the Phase 0 endpoint at epoch 3. That epoch made no weight updates (Section~\ref{sec:method-phase1}), so its points coincide with the Phase 0 endpoint and the curves first move at epoch 4.

F5 (jazz only) loses 1.4 pop points in its first updating epoch (epoch 4, at the applied learning rate of $2.56 \times 10^{-4}$), 0.67 of them from the genre marker, which it predicts as jazz from that epoch on (Section~\ref{sec:method-eval}), and continues to slide to 2.1 points below baseline by the final epoch (0.72 and 1.48 without the header targets, Table~\ref{tab:headerfree}). The onset is \emph{immediate} once updates begin, matching the catastrophic-forgetting onset described by~\cite{goodfellow2013empirical}. One updating pass over jazz-only data at that rate moves the parameters out of the pop-fluent region, and subsequent epochs deepen the loss rather than recover.

F4 (1K mix) declines more gradually, sitting 1.2 points below baseline at its best epoch and 1.5 points below by the final epoch, and 0.67 points of that is the jazz marker it predicts from epoch~4 on. Without the header targets F4 follows the same curve as F1 to F3, from $-0.17$ at epoch~4 to $-0.88$ at epoch~9, losing about 0.18 points per epoch over its first three updating epochs against 0.45 for F5, consistent with the rehearsal buffer absorbing some of the gradient that would otherwise move the model away from pop.

F1, F2, and F3 stay within a point of baseline throughout fine-tune (worst per-epoch dips $-0.71$, $-0.82$, and $-0.82$ respectively, and $-0.73$, $-0.84$, and $-0.90$ without the header targets). F1's best epoch rises above baseline ($+0.39$) only because it predicts the C-major key, and the same holds for F3's return to baseline at its final epoch. Without the header targets both lie on the common rehearsal curve ($-0.49$ and $-0.90$), and no rehearsal volume improves on the Phase 0 endpoint at any epoch.

Jazz accuracy across all runs plateaus within four to six fine-tune epochs. The plateau height is roughly equal across the five (within 1 to 2 points of each other, and within 0.7 without the header targets), reinforcing Section~\ref{sec:results-headline}: the absolute jazz ceiling is not very sensitive to rehearsal volume. The \emph{pop axis} is where the rehearsal buffer is doing real work.

\subsection{The Pareto trade-off}
\label{sec:results-pareto}

\begin{figure}[h]
  \centering
  \includegraphics[width=0.7\linewidth]{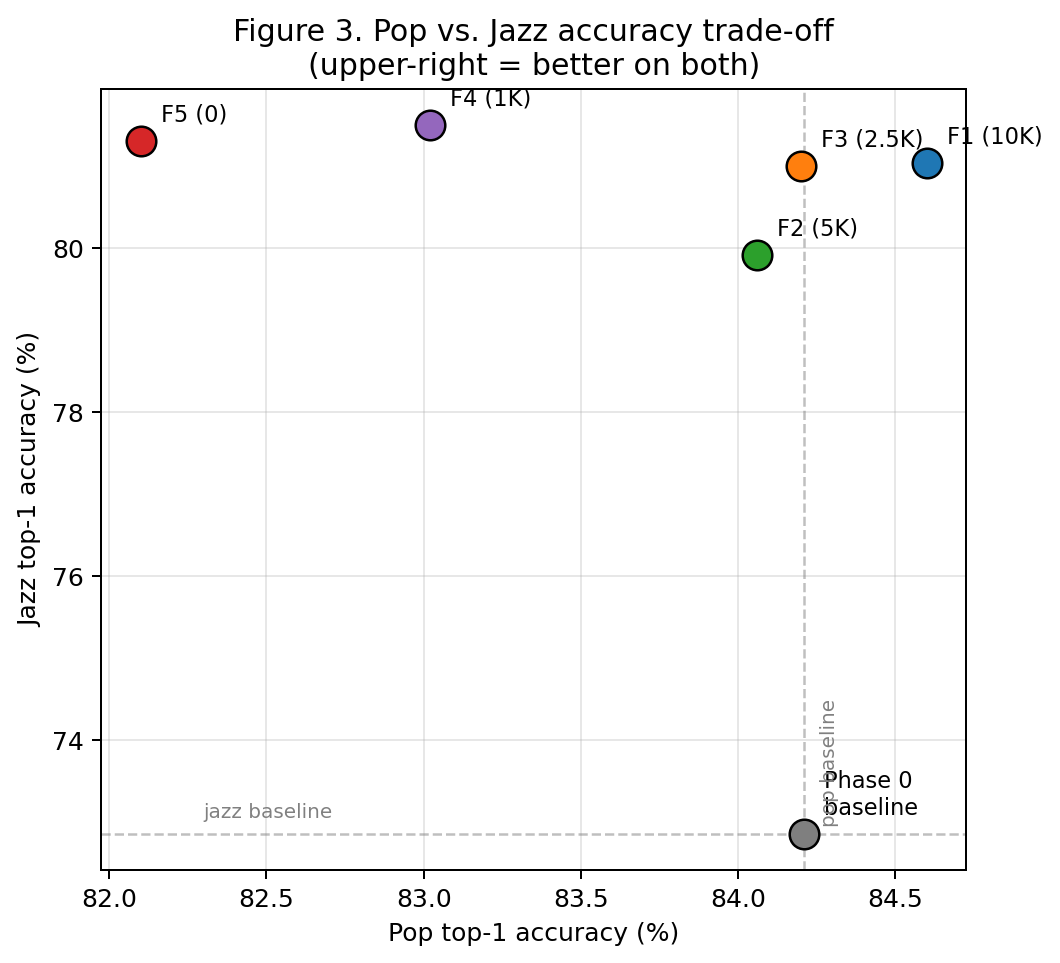}
  \caption{Pop vs.\ jazz top-1 over all targets at each run's tabulated epoch. Upper-right is best (high on both axes). On this metric F4 and F1 are the non-dominated endpoints with F3 just inside between them, while F2 and F5 are dominated, an arrangement the header targets produce (Section~\ref{sec:results-pareto}).}
  \label{fig:3}
\end{figure}

Figure~\ref{fig:3} plots the tabulated positions in (pop top-1, jazz top-1) space. Phase 0 sits in the lower-right. The five fine-tunes occupy a band in the upper region, with F1 and F3 in the upper-right, F2 just below them, and F4 and F5 further left at a slightly higher jazz top-1. On this all-target metric only F1 and F4 are non-dominated, but the arrangement comes from the header targets: F1 and F3 sit high on both axes because of the C-major key, and F4 and F5 sit low on pop and high on jazz because of the jazz marker (Section~\ref{sec:method-eval}). Without the header targets (Table~\ref{tab:headerfree}) the rehearsal runs differ mainly by tabulated epoch, later epochs losing more pop along one common curve while the jazz gain levels off from epoch~6, and F5 is dominated by F1 (same jazz, worse pop), the empirical confirmation that pure-target fine-tuning has no operating advantage here.

The frontier is clustered tightly enough that the choice among the rehearsal runs is best read as a \emph{style} decision rather than a \emph{quality} decision. Compared at the same epoch without the header targets, the token-level metric barely distinguishes F1 (87\% pop), F2 (77\%), F3 (62\%), and F4 (40\%), which lie within 0.11 points of each other on pop and within about half a point on jazz.

\subsection{Qualitative continuations}
\label{sec:results-qual}

These observations come from the author's inspection of the generated continuations and not from a controlled listening study. Section~\ref{sec:limitations} identifies the latter as the natural follow-up. The continuations below are exact, reproducible samples from the released checkpoints (deterministic seed 42; nucleus sampling at temperature 0.8 and top-p 0.9; 32 new tokens; \texttt{|} marks bar boundaries). Each is a single illustrative draw from a stochastic sampler, so the per-checkpoint differences are tendencies across many continuations rather than deterministic behaviors.

For the jazz ii--V--I prompt with input \texttt{[Dm7, G7], [Cmaj7]} in C major:

\begin{sloppypar}
\begin{itemize}
  \item \textbf{Phase 0 baseline}: \texttt{Dm7 G7 | Cmaj7 Am7 | Dm7 G7 | Cmaj7 Am7 | \ldots}. A strict diatonic ii--V--I--vi loop with no chromatic harmony.
  \item \textbf{F3 (2.5K mix)}: \texttt{Dm7 G7 | Cmaj7 Fmaj7 | Em7 A7 | Dm7 G7 | Cmaj7 | Em7 A7 | Dm7 G7 | Cmaj7 | Gm7 C7 | Fmaj7 | \ldots}. Jazz coloration enters within an otherwise diatonic frame: a secondary dominant (\texttt{A7}, V7/ii) and a secondary ii--V tonicizing the subdominant (\texttt{Gm7 C7} $\to$ \texttt{Fmaj7}).
  \item \textbf{F5 (jazz only)}: \texttt{Dm7 G7 | Cmaj7 F7 | Em7 Eb7 | Dm7 G7 | Cmaj7 | Em7b5 A7 | \ldots}. The leanest-rehearsal model reaches for denser chromatic color: a chromatic \texttt{F7} (IV7), an \texttt{Eb7} tritone-substitute dominant resolving to \texttt{Dm7}, and an applied minor ii--V of ii (\texttt{Em7b5}--\texttt{A7}, tonicizing \texttt{Dm} but left unresolved). This is the dense-chromatic register Sections~\ref{sec:discussion-endpoints} and~\ref{sec:discussion-impl} associate with the jazz-only setting.
\end{itemize}
\end{sloppypar}

For the pop I--vi--ii--V prompt every checkpoint reproduces the \texttt{Cmaj Am | Dm Gmaj | \ldots} loop verbatim, with no chromatic departures in any sample. Pop fluency on this kind of prompt is essentially saturated.

For the jazz minor ii--V prompt with input \texttt{[Bm7b5, E7], [Am7]} in A minor, the pop-leaning and jazz-leaning ends diverge most clearly. F1 (here the selection-corrected retrain \texttt{ft-pop80-v2}, since the released F1 is weight-identical to Phase~0, as noted in the Correction) keeps entirely to closely related keys (\texttt{Bm7b5 E7 | Am7 D7 | Bm7b5 E7 | Am7 D7 | Dm7 G7 | Cmaj7 Am7 | \ldots}): it circles the home-key minor ii--V--i (\texttt{Bm7b5 E7} $\to$ \texttt{Am7}, tagged with an unresolved secondary-dominant \texttt{D7}), then cadences into the relative major C (\texttt{Dm7 G7} $\to$ \texttt{Cmaj7}) and settles back on the \texttt{Am7} tonic, never modulating to a distant key. The result is smooth, pop-like motion. F4 instead spins out a chain of ii--V resolutions that drift sharpward from home A minor, through the closely related E minor and G major out to the remote B minor and F\# minor, before cadencing back to the home tonic in the parallel major, A major (\texttt{Bm7b5 E7 | Am7 | Bm7b5 E7 | Am7 | F\#m7b5 B7 | Em7 | Am7 D7 | Gmaj7 | Dbm7b5 F\#7 | Bm7 | Abm7b5 Db7 | F\#m7 | Bm7 E7 | Amaj7 | \ldots}), recognizable as bebop-style sequential harmonic motion. These differences are consistent with the Section~\ref{sec:discussion-endpoints} observation that F3 is \emph{intermediate} in stylistic register while F1 and F4 commit more to particular vocabularies.

\section{Discussion}
\label{sec:discussion}

\subsection{Choosing among the rehearsal runs}
\label{sec:discussion-sweet}

On the logged metric F3 (2.5K pop mix) looks like the balanced checkpoint, preserving pop within 0.01 points of baseline while gaining 8.13 points of jazz, F1 (10K) like the pop-preserving one ($+0.39$ pop), and F4 (1K) like the jazz-leaning one (the highest jazz number). Those distinctions come from the header targets (Section~\ref{sec:method-eval}). Without them the four rehearsal runs lose pop at nearly one rate, the 10K run at most 0.11 points less, and gain 7.2 to 7.8 jazz points, so the metric gives little reason to prefer one rehearsal volume from 1K to 10K over another. The epoch matters more than the volume: earlier epochs lose less pop (0.15 points at epoch~4 against about 0.5 at epoch~6) for a slightly smaller jazz gain (7.0 to 7.2 points against 7.4 to 7.8). The choice among the rehearsal runs is then a stylistic one, made by listening (Section~\ref{sec:discussion-endpoints}), with F3 at an intermediate register and F1 and F4 committing more to particular vocabularies, and any of them is a safe single-checkpoint default for an application serving a heterogeneous user base.

\textbf{Note (v2):} the \emph{released} F1 checkpoint coincides with the Phase~0 baseline: minimum-mixed-validation-loss selection retained the first fine-tune epoch, which the inherited learning rate of 0 had left identical to Phase~0 (see Correction). For a jazz-adapted F1 endpoint, use the selection-corrected retrain \texttt{ft-pop80-v2}. F5 is dominated and serves as a calibration point rather than a production checkpoint.

\subsection{Onset and prevention of catastrophic forgetting}
\label{sec:discussion-onset}

Figure~\ref{fig:2} shows that pop-to-jazz catastrophic forgetting is \emph{immediate}. The first updating epoch of jazz-only fine-tuning (F5, epoch 4) drops pop by 1.4 points (0.72 without the header targets, Table~\ref{tab:headerfree}), and further training deepens the loss to 2.1 points (1.48) rather than recovering it. The failure mode does not require many epochs to manifest. At the applied learning rate ($2.56 \times 10^{-4}$ in that epoch), a single updating pass of pure-target fine-tuning already costs about half of F5's final pop loss without the header targets, four to five times what any rehearsal run loses in that epoch. Whether the much lower configured rate would delay the onset is not tested here.

The corresponding observation on the rehearsal axis is that a modest buffer is highly effective. As little as 1{,}000 pop sequences (F4, $0.66\times$ the jazz training volume) cuts F5's pop loss by about half or more at every epoch once the header targets are set aside, and 2.5K, 5K, and 10K sequences stay within 0.11 points of F4 at every epoch, only the 10K run doing slightly better (0.04 to 0.11 points less loss from epoch~5 on). On the logged metric the step looks later and larger, from F4's $-1.19$ to F3's $-0.01$, because the majority genre of the mix decides which marker the model predicts and F3's tabulated epoch predicts the C-major key. No buffer stops the slow decline, which reaches 0.73 to 0.90 points by the last epoch. Any adaptation pipeline targeting a small new genre from a large pretrain should include source-genre rehearsal, and in this setting two thirds of the target-genre volume already gave nearly all the protection that any tested volume gave.

\subsection{Saturation in pop rehearsal}
\label{sec:discussion-saturation}

Without the header targets F3 (2.5K), F2 (5K), and F1 (10K) do not outperform F4 (1K) by more than 0.11 pop points at any epoch, and they match it on jazz within about half a point. The pop signal required to anchor the optimizer is small relative to the total pretrain distribution. Once the optimizer sees enough pop per epoch to regularize the update direction, additional pop adds little. Within this setting the critical rehearsal volume lies between zero and $0.66\times$ the target-genre training data, at or below the smallest nonzero volume tested. Whether this generalizes to other genre pairs, model sizes, or domains beyond chord generation is an open question. The sweep is not fine enough below 1K to settle it. For this task and architecture, the marginal value beyond 1K is small.

\subsection{Stylistic identity at the endpoints}
\label{sec:discussion-endpoints}

The token-level metric does not separate the rehearsal runs (Section~\ref{sec:discussion-sweet}), but a complementary picture emerges when one reads the \emph{endpoints} as stylistic settings rather than compromises. Each end of the rehearsal sweep carries a more distinctive harmonic identity than the middle, and the difference is audible in the Section~\ref{sec:results-qual} continuations.

F1 (pop-leaning) generates output anchored in commercial pop and rock harmony, admitting jazz coloration only selectively, for example an occasional ii--V detour or secondary dominant inside an otherwise diatonic loop. The model behaves as a pop chord generator that \emph{knows} jazz but does not commit to speaking it. F4 (jazz-leaning) leans into jazz vocabulary, including secondary dominants, tritone substitutions, modal interchange, and II--V chains across distant keys, with sparser concession to pop. The model behaves as a jazz chord generator that retains just enough pop fluency to remain stable. F3 sits between the two. It is fluent on both genres, but its outputs do not commit as strongly to either register.

In informal listening by the author across many continuations, F1 and F4 are the two checkpoints most frequently \emph{preferred} for finished compositions. F1 is preferred when the target is a familiar pop progression with one or two colorful jazz pivots. F4 is preferred when the goal is unmistakably jazz-flavored. F3 is selected less often as the \emph{preferred} output. I treat this as a hint, not a finding, since it rests on a single listener's judgment. It is consistent with a broader pattern across symbolic music generation: token-prediction accuracy on a held-out test set is not the same quantity as the perceptual quality of free continuations~\cite{briot2020deep,ji2020comprehensive}, particularly in genres like jazz where the harmonic vocabulary is small but the \emph{acceptable arrangement} of that vocabulary is large.

\subsection{Implications for chord-composition tools and the prior pipeline}
\label{sec:discussion-impl}

Section~\ref{sec:discussion-endpoints} leads to a concrete suggestion for chord-composition applications consuming models like the ones here. Rather than serving F3 alone, expose F1, F3, and F4 as user-selectable models with brief stylistic descriptions (``pop-leaning'', ``balanced'', ``jazz-leaning'') and default to F3 only because some default has to be chosen. F2 and F5 can remain available for completeness but need not be promoted, and Phase 0 can be exposed as the unmodified pop reference. The three-checkpoint surface costs almost nothing. Model-selection UI is cheap, inference is fast at this scale, and users gain a stylistic dial that the token-level metric alone would not provide. The choice is consistent with prior co-creative interaction work~\cite{huang2016chordripple,pachet2017continuator} that frames the system's role as widening the user's stylistic option space rather than autocompleting toward a single point.

The author's earlier internal pipeline was a pop pretrain followed by a jazz-only fine-tune (essentially F5 here) and was perceived by knowledgeable users as commercially unusable due to overly dense chromatic harmony. The empirical anchor for this paper came from listening to that earlier output and recognizing the pattern as catastrophic forgetting. Any of the rehearsal runs F1 to F4 addresses the failure mode, and together they give users a finer dial than the prior all-or-nothing setup. Without the empirical comparison against rehearsal-augmented runs, characterizing the failure beyond ``jazz is too dense'' would have been difficult, and there would have been no principled remediation other than reverting to the Phase 0 pretrain alone.

\section{Limitations}
\label{sec:limitations}

The setup is narrower than the most general formulation of cross-genre adaptation in symbolic music modeling.

First, I examine a single architectural family (Music Transformer with relative-position attention) at a single model size (approximately 25M parameters), with one pretraining configuration and a single fine-tune run per setting. Scaling effects (whether the same critical rehearsal ratio holds at 100M or 1B parameters or for different architectures) are not characterized.

Second, only one seed is used per configuration, and the header targets add a run-level component of their own (Section~\ref{sec:method-eval}). Without them the rehearsal runs differ by at most 0.11 pop points at any epoch, and the 10K run loses the least pop at every epoch, 0.09 to 0.11 points less than the other three by epoch~8. That gap is several times the training-seed spread of pop top-1 in the three seed retrains of the F1 recipe (at most 0.02 points at any shared epoch, on the expanded corpus of the v2 Correction), but those retrains share one rehearsal subsample, so the variation from drawing another subsample is not measured. The sweep therefore supports the contrast between no rehearsal and some rehearsal and at most a small edge for the largest volume, and it cannot place the threshold below 1K rehearsal sequences or rank the three smaller volumes.

Third, the jazz corpus is small (approximately 1{,}500 train sequences) and biased toward jazz standards, the great American songbook, and the early-to-mid bebop tradition that dominates the source corpora. Transfer to free jazz, contemporary jazz with different harmonic conventions, or other small-data harmonic styles (e.g.\ Brazilian choro, non-diatonic progressive rock) is unverified. There is also a deeper structural reason, raised in Section~\ref{sec:intro}, that limits how much of jazz any chord-symbol model can hope to capture. Jazz lives substantially in live performance, and the recorded artifact under-samples it. A chord-symbol model can faithfully reproduce the kinds of \emph{progressions} that appear on lead sheets. It cannot reproduce the moment-to-moment improvisation that defines jazz as a practice. I make no claim about that larger problem.

Fourth, \textbf{evaluation is token-level}. Top-1 and top-5 next-token accuracy on held-out chord sequences, which also scores bar, end, and header tokens (Section~\ref{sec:method-eval}), is an indirect measure of musical quality. The qualitative observations in Sections~\ref{sec:results-qual} and~\ref{sec:discussion-endpoints} partially fill this gap but rest on a single listener's judgment. \textbf{No formal listening study with multiple raters has been conducted in this version of the paper.} A controlled study comparing continuations from F1, F3, F4, and Phase 0 across genre-balanced prompts would substantially strengthen the metric-versus-aesthetic claim of Section~\ref{sec:discussion-endpoints}. I treat such a study as the natural next experiment, and a follow-up version of this paper integrating it is planned.

Fifth, the ``preferred for finished compositions'' observation in Section~\ref{sec:discussion-endpoints} reflects a single composer's workflow. Users with different goals (for example, jazz pedagogues looking for textbook-correct ii--V--I voicings) might consistently prefer F3 over either endpoint. The claim is narrow: the middle checkpoint of the sweep is not the unique checkpoint worth surfacing in a chord-composition tool, and exposing more than one is a design lever worth pulling.

\section{Conclusion}
\label{sec:conclusion}

I sweep the volume of pop ``rehearsal'' data mixed alongside fixed jazz training data across five settings, zero plus an order-of-magnitude span of nonzero volumes (1K to 10K), in pop-to-jazz chord fine-tuning. The two main findings are quantitative, and both are stated without the header tokens that top-1 also scores (Section~\ref{sec:method-eval}). Jazz capability is acquired across all five rehearsal volumes (a 7.2- to 7.8-point top-1 gain, 7 to 9 points over all targets). Every rehearsal volume tested, from $0.66\times$ the target-genre training volume up, cuts the pop forgetting of jazz-only fine-tuning by about half or more, larger volumes add at most about 0.1 point, and no volume stops a slow decline over epochs. A complementary, more tentative observation is that the middle of the sweep is not the perceptually preferred checkpoint. The endpoints carry more committed stylistic identities that the author more often selects as finished output in informal listening.

The natural follow-up is a controlled listening study with multiple raters that characterizes the metric-versus-aesthetic divergence at the endpoints. That study is left to future work and is the basis for an extended version of this paper that I anticipate submitting to a music-information-retrieval venue. I expect the empirical rehearsal ratio reported here, at most $0.66\times$ the target-genre volume, to be useful as a baseline for future work on small-genre adaptation in symbolic music modeling. Every run used the same 1{,}513 jazz sequences, so the sweep cannot tell whether the rehearsal needed scales with the target-genre volume or is a fixed count, and carrying the multiplier to target genres smaller than jazz is untested.

I close on a note connected to Section~\ref{sec:intro}. Keeping the modeling scope at the level of chord symbols was practical and principled. Jazz cannot be fully captured by any model that learns only from recorded artifacts, because the practice itself is largely unrecorded. Working at the level of symbols that musicians already share publicly through lead sheets is what makes a study like this honest about what it does and does not model. I hope the rest of the symbolic music community continues to draw that line carefully.

\subsection*{Reproducibility}

All six trained checkpoints and the selection-corrected retrain \texttt{ft-pop80-v2} are released at \url{https://huggingface.co/PearlLeeStudio}. Each model repository holds the run's per-epoch \texttt{eval\_results.csv}, the \texttt{config.json} it was launched with, the tokenizer, and the model code. The source datasets are named in the model cards, and their download links are listed in the Datasets section of the public code repository. The split and the pop rehearsal subsample use seed 42, while the training runs set no explicit random seed, so a rerun reproduces the per-epoch metrics up to seed variation. Training also used fp16 mixed precision on the RTX~4070 Mobile, so even rescoring a released checkpoint reproduces the reported values only up to a small numeric tolerance rather than bit-for-bit. Note (v2): the released \texttt{best.pt} checkpoints were selected by minimum mixed-validation loss and, for F1, F3, and F4, differ from the best-epoch metrics in Tables~4--5: the best-epoch weights for those runs were not uploaded, and the released F1 coincides with Phase~0 (see Correction). The licensed source datasets themselves are not redistributed. The code that produced the experiments is public at \url{https://github.com/PearlLeeStudio/TheArtist} (MIT license for the code, while the released weights remain CC~BY-NC~4.0).

\bibliographystyle{plain}
\bibliography{refs}

\end{document}